\documentclass[singlecol]{epl2}
\def\ds{\displaystyle}
\title{Dynamical Control in a Quasi-periodically Modulated Optical Lattice}

\author{C. Yuce}
\institute{ Department of Physics, Anadolu University, Eskisehir,
Turkey.\\cyuce@anadolu.edu.tr}

\pacs{03.75.Lm,03.75.Kk}{}

\abstract{ We investigate quantum tunneling phenomena for an
optical lattice subjected to a bichromatic $ac$ force. We show
that incommensurability of the frequencies leads to super Bloch
oscillation. We propose directed super Bloch oscillation for the
quasi periodically driven optical lattice. We study the dynamical
localization and photon assisted tunneling for a periodical and
quasi-periodical $ac$ force.}
\begin{document}

\maketitle

\section{Introduction}

Quantum tunneling of particles between potential wells is a
fundamental concepts in physics and used to perform measurements
of forces with high sensitivity and accuracy
\cite{interferometer1}. Recent experiments on cold atoms in
optical lattices subject to time-periodic perturbations have
revealed details of tunneling control through an external driving
field. Tunneling control in optical lattices depends in general on
the form of the external potential and, in particular, on its
parameters. Nontrivial dynamical effects have been demonstrated
even in the absence of nonlinear atom-atom interaction. In an
optical lattice, the Bloch states are completely delocalized, with
the atomic wave functions extending over the whole lattice. Thus,
an initially localized wave-packet spreads ballistically in time.
The presence of $dc$ force breaks the translational symmetry and
suppress atomic tunneling \cite{blochosc1,blochosc2}. The physical
picture changes if the external force is $ac$ type (optical
lattice are periodically shaken back and forth). The effect of an
$ac$ force is the reduced tunneling rate. At certain shaking
strengths, tunneling is lost and dynamical localization occurs
\cite{dunlap,dynloc1}. The combined presence of both $dc$ and $ac$
forces leads interesting results such as directed motion, super
Bloch oscillation and photon-assisted tunneling. Super Bloch
oscillations over hundreds of lattice sites occur when an integer
multiple of the $ac$ frequency is slightly detuned from the Bloch
frequency associated with the $dc$ component of the force
\cite{SBloch1,SBloch2,SBloch21,SBloch22,SBlochyeni,thommen,1}. In
photon assisted tunneling, wavepacket delocalization occurs for an
initially localized wave packet when modulation frequency is at
multiple integers of the Bloch frequency
\cite{Passist1,Passist2,Passist3,Passist4,Passist5,onemli,tino}.
It was shown that not only the strength of the $ac$ force but also
its phase is of experimental relevance \cite{phase,phase2}. The
generalization of the monochromatic $ac$ force to a
multi-frequency $ac$ force leads to nontrivial results in ultra
cold atom dynamics
\cite{bichromatic0,bichromatic1,bichromaticonemli1,bichromaticonemli1devamý,bichromaticonemli3,bichromatic2,bichromatic3,bichromatic4,bichromatic5,bichromatic6,bichromatic7,bichromaticrathced1}.
The relationship between symmetries and transport was explored for
a cold atom ratchet with multifrequency driving
\cite{bichromaticonemli1devamý}. It was shown that the bi-harmonic
$ac$ force stabilizes the dynamics, allowing the generation of
uniform directed motion over a range of momentum much larger than
what is possible with a $dc$ bias \cite{bichromatic4}. \\
In this paper, we study quantum tunneling phenomena for an optical
lattice subjected to a bichromatic $ac$ force. We obseve that
nontrivial tunneling phenomenon appears for the system under
consideration. We show that incommensurability of the frequencies
and phases of the bichromatic $ac$ force play important roles on
the dynamical localization and photon assisted tunneling. We
discuss that super Bloch oscillation occurs if the bichromatic
$ac$ force is quasi-periodical. We propose directed super Bloch
oscillation.

\section{Bichromatic Field}

\begin{figure}[htb]
\begin{center}
\centering
\includegraphics[width=6cm]{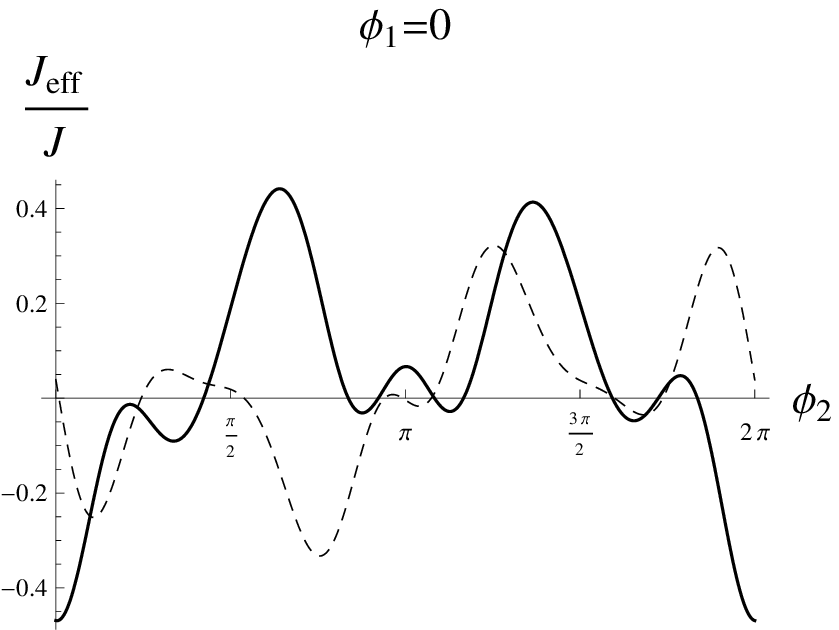}
\includegraphics[width=6cm]{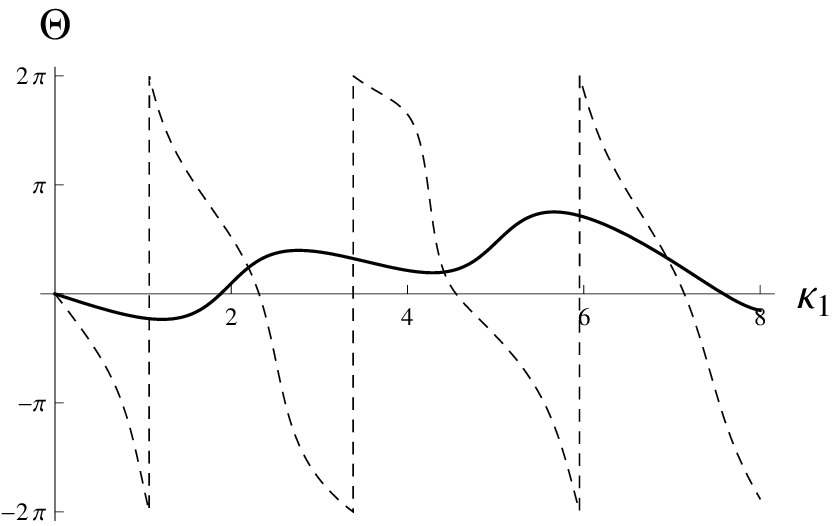}
\caption{1-) The real (solid) and imaginary (dashed) parts $\ds{
J_{eff}/J}$ at $\ds{n_f=0}$, $\ds{\gamma=1/3}$ and
$\ds{\kappa_2=\kappa_1=1}$ versus the phases. 2-) The argument
$\ds{\Theta}$ of the effective tunneling at $\ds{\gamma=1/2}$
(dashed) and $\ds{\gamma=2}$ (solid) versus
$\ds{\kappa_1=\kappa_2}$. We take $\ds{\phi_1=0,\phi_2=\pi/3}$ and
$n_f=0$. } \label{phases}
\end{center}
\end{figure}

Consider a one-dimensional driven optical lattice described by a
time-dependent single-particle Hamiltonian in the tight-binding
regime
\begin{eqnarray}\label{ham}
H=-J \sum_l (|l+1><l|+|l><l+1|)+F(t)a\sum_ll|l><l|
\end{eqnarray}
where $J$ is the tunneling rate, $a$ is the lattice constant and
$|l>$ denotes a Wannier state localized at the $l$-th lattice
site. Our system is subjected to a constant $dc$ force plus a
bichromatic $ac$ force of frequencies $\omega_0$ and
$\gamma\omega_0$, where the parameter $\ds{\gamma}$ is the ratio
of frequencies.
\begin{eqnarray}\label{mod2}
F(t)=\frac{\hbar\omega_0}{a}\left(n_f+\kappa_1
\cos({{\omega_0}t+\phi_1}) +\kappa_2
\cos({\gamma{\omega_0}t+\phi_2})\right)
\end{eqnarray}
where $\ds{\phi_1}$ and $\ds{\phi_2}$ are the phases, $\ds{n_f}$
is an integer and the dimensionless constants $\ds{\kappa_1}$ and
$\ds{\kappa_2}$ describe the strengths of the $ac$ force in units
of $\hbar\omega_0/a$. Such a force can be experimentally realized
by phase-modulating of the lattice beams
\cite{bichromaticonemli1}. Suppose that frequencies and the
strengths are small enough that the dynamics is restricted to the
lowest band. The frequencies are commensurable if $\ds{\gamma}$ is
a rational number and incommensurable if it is an irrational
number. Note that $\ds{\gamma}$ can be given with a finite number
of digits in a real experiment. To increase the incommensurability
of $\ds{\gamma=p/q}$ ($p$, $q$ are two coprime positive integers),
one can choose sufficiently large $p$ and $q$. Then the driving
force becomes effectively quasiperiodic on the time scale of the
experiment. Another possible way to get an irrational number is to
approximate it with a rational number by using the continued
fraction expansion
$\ds{\gamma=\frac{1}{a_1+\frac{1}{a_2+\frac{1}{a_3+...}}}}$, where
$\ds{a_i}$ are integers. By truncating, one increase the actual
\textit{degree of commensurability} \cite{ekref}. More precisely,
$\gamma$ acts as if it is irrational, if $\gamma T$ is not an
integer, where $T$ is the duration of the force.\\
The presence of a driving force corresponds to a modification of
the tunneling parameter. At times $\ds{t>>1/\omega_0}$ and
$\ds{t>>\gamma/\omega_0}$, the tunneling parameter $J$ is replaced
by an effective tunneling parameter, $|J_{eff.}| < |J|$ and the
Hamiltonian (\ref{ham}) can effectively be described as
$\ds{H_{eff.}=- \sum_l
J_{eff.}|l+1><l|+J_{eff.}^{\star}|l><l+1|}$. The calculation of
$\ds{J_{eff.}}$ can be found in the Appendix. It reads
\begin{eqnarray}\label{effdynloc}
\frac{J_{eff.}}{J}= e^{-i\Phi_{0}}\left( \mathcal{J}_{-n_f}
(\kappa_1)\mathcal{J}_{0}
(\frac{\kappa_2}{\gamma})+\sum_{m{\neq}0}e^{im\Phi_{\gamma}}\mathcal{J}_{-n_f-m\gamma}
(\kappa_1)\mathcal{J}_{m} (\frac{\kappa_2}{\gamma})\right)
\end{eqnarray}
where
$\ds{\Phi_0=\kappa_1\sin\phi_1+\frac{\kappa_2}{\gamma}\sin\phi_2
}$ is a phase which affects the effective tunneling as a whole,
$\ds{\Phi_{\gamma}=\phi_2-\gamma\phi_1}$ and $\mathcal{J}_m$ is
the $m$-th order Bessel function of first kind. There are
infinitely many terms in the summation. However, only a few terms
are practically important for small values of $\kappa$ since the
Bessel functions decrease with $m$: $\ds{\mathcal{J}_{m-1}
(\kappa)/\mathcal{J}_{m} (\kappa)\approx2m/\kappa}$. Of special
importance is the case without $dc$-force. Note that $\ds{
\mathcal{J}_{-n_f} (0)}$ is zero unless $n_f=0$. Hence if the
force is monochromatic, $\kappa_1=0$, the effective tunneling does
not vanish only if the $dc$ force is absent, $n_f=0$. \\
The phase $\ds{\Phi_0}$ can be eliminated by shifting the time
origin. It was discussed in \cite{phase,phase2} that the phase
$\ds{\Phi_0}$ can be of experimental relevance if the driving
potential is assumed to be switched on at $\ds{t=0}$. It would be
irrelevant if we assume that the driving is switched on at
$t\rightarrow-\infty$. However, the phase $\ds{\Phi_{\gamma}}$
inside the summation can not be eliminated even if the driving is
assumed to be switched on at $t\rightarrow-\infty$. Observe that
$\ds{|J_{eff.}|}$ does not depend on the phases $\ds{\phi_1,
\phi_2}$ only if the $ac$ force is quasi-periodic. This is because
the summation is always zero irrespective of $\ds{\kappa_1}$ and
$\ds{\kappa_2}$ when the force is quasi-periodic in time, i.e.
$\ds{\gamma}$ is an irrational number (there exists no integer $m$
such that $\ds{m\gamma}$ is an integer). We conclude that the
effective tunneling vanishes in the presence of a quasi-periodic
force if either $\kappa_1$ or $\kappa_2/\gamma$ are the roots of
the Bessel function of order $n_f$ or zero, respectively. Let us
now study briefly the case of periodic $ac$ force. In this case,
the phases $\ds{\phi_1, \phi_2}$ can be used to control the real
and imaginary parts of effective tunneling energy as can be seen
in the first figure of the Fig-(\ref{phases}), where we plot the
real and imaginary parts of $\ds{J_{eff}/J}$ versus the phase
$\ds{\phi_2}$ at $\ds{\phi_1=0}$. It was shown in \cite{phase2}
that expansion and center of mass motion of an initially Gaussian
wave packet are directly related to the real and imaginary parts
respectively of the effective tunneling. Therefore the wave packet
moves without changing the shape when $\ds{J_{eff}}$ is purely
imaginary and it expands without translating when $\ds{J_{eff}}$
is purely real. More specifically, the complex effective tunneling
(\ref{effdynloc}) can be rewritten as
$\ds{J_{eff.}=|J_{eff.}|~e^{i \Theta}}$, where $\ds{\Theta}$ is
known as  Peierls phase, which has recently been experimentally
demonstrated to realize artificial magnetic fields for neutral
atoms trapped in a one-dimensional optical lattice
\cite{peierls1,peierls2}. The effect of a Peierls phase is to
shift the minimum of the band structure to a quasimomentum
$\ds{k_{min} =\Theta/d}$. The group velocity vanishes when
$\ds{\Theta}$ is an integer multiple of $\ds{\pi}$ and is maximum
when $\ds{\Theta}$ is either $\ds{\pi/2}$ or $\ds{3\pi/2}$. Note
that the group velocity is reversed when $\ds{ \Theta=3\pi/2}$.
This leads to the concept of the directed motion. In
Fig-{\ref{phases}}, we plot the Peierls phase as a function of
$\ds{\kappa_1}$ when $\ds{n_f=0}$, $\ds{\phi_1=0}$,
$\ds{\phi_2=\pi/3}$ and $\ds{\kappa_2=\kappa_1}$.\\
The expression for the effective tunneling allows us to understand
the tunneling and localization. Below, we will study dynamical
localization, super Bloch oscillation and photon assisted
tunneling in detail.

\subsection{Dynamical Localization}

\begin{figure}[htb]
\begin{center}
\centering
\includegraphics[width=6cm]{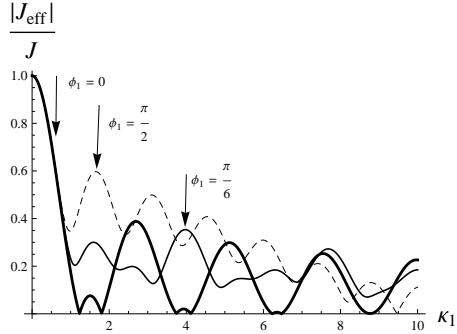}
\caption{ $\ds{|J_{eff.}|/J}$ versus $\kappa_1$ when
$\ds{\kappa_2=\kappa_1}$ for three different values of
$\ds{\phi_1}$ at fixed $\ds{n_f=0}$, $\ds{\phi_2=0}$ and
$\ds{\gamma=1/2}$.} \label{cem}
\end{center}
\end{figure}

Let $\ds{\mathcal{R}\{J_{eff.}\}}$ and
$\ds{\mathcal{I}\mathcal{M}\{J_{eff.}\}}$ be the real and
imaginary parts of the effective tunneling, respectively. If the
real part of $J_{eff.}$ is different from zero, then an initially
localized wave packet spreads in time and gets delocalized.
Suppose that $\ds{\mathcal{R}\{J_{eff.}\}=0}$. If
$\ds{\mathcal{I}\mathcal{M}\{J_{eff.}\}}$ is zero, too, dynamical
localization occurs. If
$\ds{\mathcal{I}\mathcal{M}\{J_{eff.}\}\neq0}$, then an initially
localized wave packet is still non-spreading but moves with a
constant group velocity. Here, we study dynamical localization for
the quasi periodically driven optical lattice. We consider that
both real and imaginary parts of effective tunneling
are zero.\\
For the monochromatic $ac$ force, $\kappa_2=0$, $\ds{|J_{eff.}|=
\mathcal{J}_{-n_f} (\kappa_1)~J}$. Hence, the particle is
effectively localized whenever $\ds{\kappa_1}$ is a root of the
Bessel function of order $n_f$. For the bichromatic $ac$ force,
not only $\ds{\kappa_1}$, but also the other parameters in the
Hamiltonian can be used to suppress tunneling. As discussed above,
the summation in (\ref{effdynloc}) vanishes if $\ds{\gamma}$ is an
irrational number. Hence we conclude that dynamical localization
occurs whenever either $\kappa_1$ or $\kappa_2/\gamma$ are the
roots of the Bessel function of order $\ds{n_f}$ or zero,
respectively when $\ds{\gamma}$ is an irrational number. The
phases $\ds{\phi_1}$ and $\ds{\phi_2}$ play no role on dynamical
localization if the force is quasi-periodical. However, for a
periodical $ac$ force, the summation is in general nonzero and
varying the phases changes the values of $\ds{\kappa_1}$ and
$\ds{\kappa_2}$ for the onset of dynamical localization. This
doesn't necessarily mean that one can always find a set of
parameters in the Hamiltonian so that dynamical localization
occurs. As an example, we fix $\ds{\phi_2=0}$ and vary
$\ds{\phi_1}$ for commensurate frequencies with $\gamma=1/2$. In
Fig-{\ref{cem}}, we plot the ratio $\ds{|J_{eff.}|/J}$ versus
$\kappa_1$ when $\ds{\kappa_2=\kappa_1}$. The effective tunneling
oscillates with a decaying consecutive peaks as $\kappa_1$
increases. One can see that $\ds{J_{eff.}}$ vanishes at some
particular strength $\kappa_1$ when $\ds{\phi_1=0}$. However,
dynamical localization does not occur at any $\kappa_1$ when
$\ds{\phi_1=\pi/6}$. We emphasize that the
next-to-nearest-neighbor tunneling matrix element in a tight
binding optical lattice is still nonvanishing. Thus dynamical
localization for the optical lattice driven with the force
(\ref{mod2}) is not
exact but appreciable.\\

\subsection{Super Bloch Oscillation}

A giant matter-wave oscillation that extend over hundreds of
lattice sites was observed experimentally in \cite{SBloch22}. This
large amplitude oscillation is called super Bloch oscillation. It
arises in an optical lattice driven by a $dc$ plus an $ac$ force
provided that an integer multiple of the $ac$ frequency is only
slightly detuned from the Bloch frequency associated with the $dc$
component of the force. In this section, we will show that super
Bloch oscillation can also be observed in an optical lattice
driven by a quasi periodical $ac$ force. The prediction can be
tested through experiments. We will also study the effect of an additional off-resonant $dc$ force.\\
To observe super Bloch oscillation, it is necessary to require
that the real part of effective tunneling must vanish,
$\ds{\mathcal{R}\{J_{eff.}\}=0}$. Note that vanishing
$\ds{\mathcal{R}\{J_{eff.}\}}$ does not mean entire destruction of
tunneling because of the neglected off-resonant terms in the
derivation of the effective tunneling. Our approximation is true
only when $t>>1/\omega_0$ and $t>>\gamma/\omega_0$. In fact, the
particles are not frozen but make small amplitude oscillations.
Our aim is to find a way how we can increase the amplitude of such
oscillation. Suppose $\ds{\mathcal{I}\mathcal{M}\{J_{eff.}\}=0}$
so that the center of the oscillation is fixed. If
$\ds{\mathcal{I}\mathcal{M}\{J_{eff.}\}{\neq}0}$, then the wave
packet translates when oscillating. We call this effect
\textit{directed super Bloch oscillation}. It is worth studying it
in a 2-D optical lattice. One can control 2-D motion and obtain
different trajectories \cite{thommen2}.\\
Suppose that $\ds{\gamma}$ is an irrational number and either
$\kappa_1$ or $\kappa_2/\gamma$ are the roots of Bessel function
of order $n_f$ or zero, respectively (hence$\ds{ J_{eff.}{=}0}$).
Let us choose an irrational $\ds{\gamma}$ in the very small
neighborhood of $\ds{\gamma_r}$:
$\ds{\gamma=\gamma_r+\Delta\gamma}$ where $\ds{\gamma_{r}}$ is a
rational number and $\ds{\Delta\gamma}$ is an irrational number
such that $\ds{\Delta\gamma<<1}$. We also demand that the Bloch
frequency associated with the $dc$ force is not an integer
multiple of $\omega_0$ and so we suppose
$\ds{n_f{\rightarrow}n_f+\delta_f}$, where $\ds{\delta_f}$ is a
small detuning, $\ds{\delta_f<<1}$. Therefore, the mean
displacement is given by (see the Appendix for details)
\begin{eqnarray}\label{sdkdka}
\overline{x}_g=2aJ\sum_{m}\mathcal{J}_{-n_f-m\gamma_r}
(\kappa_1)\mathcal{J}_{m} (\frac{\kappa_2}{\gamma})~
\frac{\cos({(m\Delta\gamma+\delta_f)\omega_0t+\Phi_{m}})-\cos(\Phi_{m})}{\omega_0(m\Delta\gamma+\delta_f)}
\end{eqnarray}
where $\ds{\Phi_m=m\Phi_{\gamma_r} - \Phi_0 }$  and
$\ds{\Phi_{\gamma_r}}$ and $\ds{\Phi_0}$ are given below the Eq.
(\ref{effdynloc}). As a special case of
$\ds{\kappa_2=\gamma_r=\Delta\gamma=0}$, it reduces to the well
known formula for super Bloch oscillation in the presence of $dc$
plus monochromatic $ac$ force. In this case $n_f$ must be an
integer. Observe that $n_f$ needn't be restricted to integer if
the $ac$ force is bichromatic. The constant $\ds{n_f}$ is chosen
to be equal to an integer plus
an integer multiple of $\ds{\gamma_r}$.\\
The resonance leading to ballistic expansion occurs when the
denominator in (\ref{sdkdka}) is zero. In the absence of $dc$
force, the resonance at $\ds{m=0}$ would lead to ballistic
expansion in time, if either $\kappa_1$ or $\kappa_2/\gamma$ are
not supposed to be the roots of the corresponding Bessel
functions. The detuning of the additional $dc$ force removes the
degeneracy at $m=0$. The resonance occurs only if $\ds{\delta_f}$
is an integer multiple of $\Delta\gamma$. In this case, in order
to cancel the resonance, we require either $\kappa_1$ or
$\kappa_2/\gamma$ is a root of Bessel function of order
$\ds{n_f+\gamma_r\delta_f/\Delta\gamma}$ and
$\ds{\delta_f/\Delta\gamma}$, respectively.\\
Observe that the oscillation amplitude scales as
$\ds{1/\Delta\gamma}$ when $\ds{\delta_f=0}$ and $\ds{1/\delta_f}$
when $\ds{\Delta\gamma=0}$. If we choose $\ds{\Delta\gamma<<1}$
when $\ds{\delta_f=0}$ (or $\ds{\delta_f<<1}$ when
$\ds{\Delta\gamma=0}$), the amplitude of the oscillation can be
enhanced and giant amplitude oscillation can be observed. The
amplitude of the oscillation depend strongly on $\ds{\gamma_{r}}$
since it determines the
orders of the Bessel functions in the summation. \\
The oscillation (\ref{sdkdka}) can be thought of the addition of
many simple harmonic motions whose frequencies are multiples of a
fundamental frequency and whose amplitudes are given by a product
of two Bessel functions. We define the fundamental angular
frequency as $(m^{\star}\Delta\gamma+\delta_f)~\omega_0$, where
$\ds{m^{\star}}$ is the smallest positive integer such that
$\ds{m^{\star}\gamma_r}$ becomes an integer. The interfering
harmonic oscillations are not exactly in phase and the direction
of their motion depends on the sign of Bessel functions. However,
this does not necessarily mean that the net result is negligible
due to the cancelation. Only a few terms in the summation are
practically important and generally the addition results in large
amplitude oscillation. The resultant motion is not simple harmonic
but periodical since the frequencies are all harmonics of
fundamental frequency. Hence the period of $x_g$ is the same as
the period of the fundamental oscillation:
$\ds{T=\frac{2\pi}{(m^{\star}\Delta\gamma+\delta_f)\omega_0}}$.\\
So far, we have considered that imaginary part of the effective
tunneling vanishes, which implies that the center of super Bloch
oscillation is fixed. Consider now that not the imaginary part but
the real part of effective tunneling vanishes. In this case, the
equation (\ref{sdkdka}) modifies to
$\ds{x_g\rightarrow\overline{x}_g-2a\mathcal{I}\mathcal{M}\{J_{eff.}\}}$
(see the Appendix). The sign of
$\ds{\mathcal{I}\mathcal{M}\{J_{eff.}\}}$ determines the direction
of the center of the oscillation. Since we can direct the super
Bloch oscillation, we call this effect directed super Bloch
oscillation. We think interesting trajectories can be obtained in
a 2-D optical lattice.

\subsection{Photon Assisted Tunneling}

\begin{figure}[htb]
\begin{center}
\centering
\includegraphics[width=6cm]{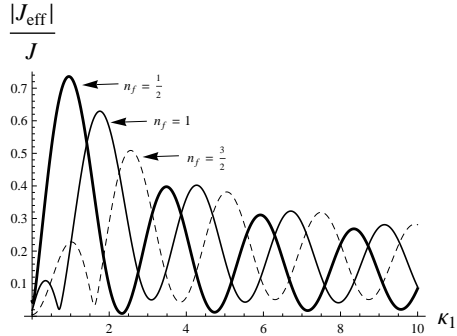}
\caption{The absolute value of $\ds{J_{eff}/J}$ versus the
$\ds{\kappa_1}$ for $\ds{n_f=\frac{1}{2},1,\frac{3}{2}}$. We take
$\ds{\kappa_2=\kappa_1}$, $\ds{\phi_1=\phi_2=0}$ and
$\ds{\gamma=1/2}$.} \label{pat}
\end{center}
\end{figure}

In an optical lattice subjected to no force, the Bloch states are
completely delocalized, with the atomic wave functions extending
over the whole lattice. Thus, an initially localized wave-packet
spreads ballistically in time. The presence of $dc$ force breaks
the translational symmetry and suppress atomic tunneling
\cite{blochosc1,blochosc2}. Thus, an initially localized
wave-packet remains localized during time evolution. More
specifically, the wave-packet width slightly oscillates during the
Bloch oscillation. The picture changes if monochromatic $ac$ force
is applied together with $dc$ force. The presence of $ac$ force
restore tunneling. This effect is known as photon assisted
tunneling \cite{Passist1}. The role of the photons is played by a
periodic shaking of the lattice at the resonant frequencies. In
this subsection, we will study photon assisted tunneling for the
$dc$ plus bichromatic $ac$ force.\\
The absolute value of the effective tunneling (\ref{effdynloc}) is
reduced to a simple formula when $\ds{\gamma}$ is an irrational
number: $\ds{|J_{eff}|= \mathcal{J}_{-n_f} (\kappa_1)
\mathcal{J}_{0} (\kappa_2/\gamma)J}$. Note that $\ds{
\mathcal{J}_{n_f} (\kappa_1)}$ is zero at $\ds{\kappa_1=0}$ unless
$n_f{=}0$. In the absence of the $ac$ force,
$\ds{\kappa_1=\kappa_2=0}$, the effective tunneling is zero.
Tunneling is restored only when $ac$ force is present in the
system. The effect of the second $ac$ force with coupling
$\kappa_2$ is the reduced effective tunneling. If the $ac$ force
is periodic in time, then the frequency associated with $dc$ force
is not restricted to an integer multiple of $\ds{\omega_0}$ for
the photon assisted tunneling effect. Tunneling is restored if
$\ds{n_f}$ is either an integer or an integer multiple of
$\ds{\gamma}$ plus an integer. The Fig-{\ref{pat}} plots the
absolute value of $\ds{J_{eff}/J}$ as a function of
$\ds{\kappa_1}$ for different values $\ds{n_f}$. In the presence
of the $ac$ force at an appropriate strength, tunneling is
partially restored. Note also that $\ds{|J_{-n_f}|=|J_{n_f}|}$
when $n_f$ is an integer. So, reversing the sign of $n_f$ change
the tunneling amplitude only when the summation is non-vanishing,
i. e. the $ac$ force is periodical.

\subsection{Discussion}

In this paper, we consider the bichromatic $ac$ force and show how
incommensurability of the frequencies leads to nontrivial physical
effects. Suppose now that quasi-periodical $ac$ force is
multicolored. As an example, consider $
F(t)=\frac{\hbar\omega_0}{a}(\sum_{n=1}\kappa_{n}
\cos({n{\omega_0}t+\phi_n}) +\kappa
\cos({\gamma{\omega_0}t+\phi}))$, where the first term is the
Fourier series expansion of a periodic function and $\gamma$ is an
irrational number such that the force is quasi-periodic. In this
case, the absolute value of the effective tunneling is given by
$\ds{|J_{eff}|/J=\mathcal{J}_{0}
(\frac{\kappa}{\gamma})\prod_{n=1}\mathcal{J}_{0}
(\frac{\kappa_n}{n})}$, where $n$ is the index of multiplication
in the product symbol. Recall that the absolute value of the
Bessel function of order zero is always less than one. Since there
are infinitely many terms in the product, the effective tunneling
is zero independent of $\ds{\kappa_n}$. It is generally believed
that dynamical localization occurs at some certain parameters in
the Hamiltonian. Here, we show that dynamical localization can
also occur regardless of the
parameters.\\
To sum up, we have considered an optical lattice driven by a
bichromatic $ac$ force. We have shown that the effective tunneling
depends sensitively on the commensurability of frequencies and the
phases. We have also shown that depending on the parameters in the
Hamiltonian, quasi-periodicity results in the suppression of
tunneling and leads to super Bloch oscillation. We propose
directed super Bloch oscillation. We have also considered
dynamical localization and photon assisted tunneling for a
periodical and quasi-periodical $ac$ force.

\section{Appendix}

Let us define $\ds{I(t)}$ as
\begin{eqnarray}\label{aas}
I(t)&=&\int_0^t \exp{i(\eta)} ~J~ dt^{\prime}~.
\end{eqnarray}
where $\eta$ is given by
$\ds{\eta(t)=\frac{a}{\hbar}\int_0^{t}{F(t^{\prime})~dt^{\prime}}}$.
The expression $I(t)$ is of importance to understand tunneling
phenomena at times $\ds{t>>1/\omega_0}$ and
$\ds{t>>\gamma/\omega_0}$. Suppose that we take a Gaussian wave
packet as an initial state. The system's wave function on lattice
site $l$ reads $\ds{\psi_l=Ne^{-l^2/2\sigma_0^2}}$ where
$\sigma_0$ is the initial width of the wavepacket measured in
units of the lattice spacing, $N$ is the normalization constant.
The wavepacket expansion and its center of mass motion are
directly related to the real and imaginary parts respectively of
$\ds{I(t)}$ \cite{phase2}: $\ds{\sigma(t)=\sigma_0 \sqrt{1+
(\mathcal{R}
\{I\}/\sigma_0^2)^2}}$; $\ds{\bar{x}_g=-2a\mathrm{I}\mathrm{m} \{I\} }$.\\
To understand the tunneling dynamics, we expand the oscillatory
term $\ds{e^{i\eta}}$ in terms of Bessel functions by using the
Jacobi-Anger expansion; $\ds{ e^{i\kappa\sin(x)}=\sum_{l}
\mathcal{J}_l ( \kappa )e^{ilx} }$. If we substitute  the
Jacobi-Anger expansion into the integration $I(t)$ for the force
(\ref{mod2}), we get
\begin{equation}\label{besselacilimi}
I(t)=Je^{-i\Phi_0}\sum_{l,m}  \mathcal{J}_l ( \kappa_1
)\mathcal{J}_m ( \frac{\kappa_2}{\gamma}  ) e^{i(l\phi_1+m\phi_2)}
S_{l,m}
\end{equation}
where $\ds{\Phi_{0}=\kappa_1\sin\phi_2+\kappa_2/\gamma\sin\phi_2
}$ and $\ds{S_{l,m}=\int_0^t e^{i(l+m\gamma)\omega_0{t^{\prime}}}
dt^{\prime} }$. It is straightforward to generalize this
expression to the case
of the additional presence of the $dc$ force.\\
At resonance $\ds{(l+m\gamma=0)}$, the integral $\ds{S_{l,m}}$
becomes unbounded as $\ds{t{\rightarrow}\infty}$. Ignoring the
off-resonant terms at large times, we get
$\ds{S_{l,m}=\delta_{-m\gamma,m}~t}$. Therefore the unbounded
solution can be given by $\ds{I=J_{eff.} t}$, where $\ds{J_{eff.}
}$ is the effective tunneling rate (\ref{effdynloc}). Note that
there is another alternate way to get $\ds{J_{eff.} }$.
The time average of $\ds{I }$ gives the effective tunneling.\\
Secondly, let us suppose $\ds{\gamma=\gamma_r+\Delta\gamma}$ where
$\ds{\gamma_{r}}$ is a rational number and $\ds{\Delta\gamma}$ is
an irrational number such that $\ds{\Delta\gamma<<1}$. We can then
study super Bloch oscillation with this choice of $\gamma$.
Therefore, the integral becomes
$\ds{S_{l,m}=\delta_{-m\gamma_r,m}\int_0^t
e^{im\Delta\gamma\omega_0{t^{\prime}}} dt^{\prime} }$. The
smallness of $\ds{\Delta\omega}$ leads to large amplitude
oscillation. Using the relation
$\ds{\bar{x}_g=-2a\mathrm{I}\mathrm{m} \{I\} }$, we get the Equ.
(\ref{sdkdka}).


\begin{thebibliography}{0}
\bibitem{interferometer1} \Name{N. Poli, F.-Y. Wang, M. G. Tarallo, A. Alberti, M. Prevedelli, G. M. Tino} \REVIEW{ Phys. Rev. Lett.} {106}{2011}{038501}.
\bibitem{blochosc1} \Name{G. Ferrari, N. Poli, F. Sorrentino, G. M. Tino} \REVIEW{ Phys. Rev. Lett.} {97} {2006} {060402}.
\bibitem{blochosc2} \Name{O. Morsch, J. H. Müller, M. Cristiani, D. Ciampini and E. Arimondo} \REVIEW{ Phys. Rev. Lett.} {87}  {2001}{140402}.
\bibitem{dunlap} \Name{D. H. Dunlap and V. M. Kenkre} \REVIEW{ Phys. Rev. B} {34} {1986} {3625}.
\bibitem{dynloc1} \Name{A. Eckardt, M. Holthaus, H. L. A. Zenesini, D. Ciampini, O. Morsch, E. Arimondo}\REVIEW{ Phys. Rev. A} {79}{2009} {013611} .
\bibitem{SBloch1} \Name{K. Kudo T.S. Monteiro} \REVIEW{ Phys. Rev. A} {83}{2011} {053627}.
\bibitem{SBloch2} \Name{Stephan Arlinghaus Martin Holthaus} \REVIEW{ Phys. Rev. B} {84} {2011}{054301}.
\bibitem{SBloch21} \Name{A. Alberti V. V. Ivanov  G. M. Tino and G. Ferrari} Nature \REVIEW{ Physics} {5}{2009} {547} .
\bibitem{SBloch22} \Name{E. Haller R. Hart M. J. Mark J. G. Danzl L. Reichsollner and H. C. Nagerl} \REVIEW{ Phys. Rev. Lett.} {104} {2010} {200403}.
\bibitem{SBlochyeni} \Name{Andrey R. Kolovsky} \REVIEW{ Phys. Rev. A} {82} {2010} {011601R}.
\bibitem{thommen} \Name{Quentin Thommen, Jean Claude Garreau and Veronique Zehnle} \REVIEW{ Phys. Rev. A} {65} {2002} {053406}.
\bibitem{1} \Name{H. Lignier et al} \REVIEW{ Phys. Rev. Lett.} {99}{2007} {220403}.
\bibitem{Passist1} \Name{C. Sias, H. Lignier, Y. P. Singh, A. Zenesini, D. Ciampini, O. Morsch and E. Arimondo} \REVIEW{ Phys. Rev. Lett.} { 100} {2008}{040404} .
\bibitem{Passist2} \Name{A. Eckardt, T. Jinasundera, C. Weiss and M. Holthaus} \REVIEW{ Phys. Rev. Lett.} {95}{2005} {200401}.
\bibitem{Passist3} \Name{Christoph Weiss, Heinz-Peter Breuer} \REVIEW{ Phys. Rev. A} {79} {2009} {023608}.
\bibitem{Passist4} \Name{Andre Eckardt, Tharanga Jinasundera, Christoph Weiss, and Martin Holthaus} \REVIEW{ Phys. Rev. Lett.} {95}  {2005}{200401}.
\bibitem{Passist5} \Name{Q. Beaufils, G. Tackmann, X. Wang, B. Pelle, S. Pelisson, P. Wolf and F. Pereira dos Santos} \REVIEW{ Phys. Rev. Lett.} {106} {2011} {213002}.
\bibitem{onemli} \Name{A. Alberti, G. Ferrari, V. V. Ivanov, M. L. Chiofalo and G. M. Tino}  \REVIEW{New J. Phys.} {12} {2010} {065037}.
\bibitem{tino} \Name{V.V. Ivanov, A. Alberti, M. Schioppo, G. Ferrari, M. Artoni, M. L. Chiofalo, G. M.Tino} \REVIEW{ Phys. Rev. Lett.} {100}{2008} {043602} .
\bibitem{phase} \Name{C. E. Creffield, F. Sols} \REVIEW{ Phys. Rev. Lett.} {100} {2008} {250402}.
\bibitem{phase2} \Name{C. E. Creffield, F. Sols} \REVIEW{ Phys. Rev. A}  {84} {2011} {023630}.
\bibitem{bichromatic0} \Name{A. Klumpp, D. Witthaut, H. J. Korsch} \REVIEW{ J. Phys. A: Math. Theor. } {40} {2007} {2299}.
\bibitem{bichromatic1} \Name{M. Schiavoni, L. Sanchez-Palencia, F.Renzoni and G. Grynberg} \REVIEW{ Phys. Rev. Lett.} {90} {2003} {094101}.
\bibitem{bichromaticonemli1} \Name{R. Gommers, S. Denisov and F. Renzoni} \REVIEW{ Phys. Rev. Lett.} {96} {2006} {240604}.
\bibitem{bichromaticonemli1devamý} \Name{R. Gommers, M. Brown and F. Renzoni} \REVIEW{ Phys. Rev. A} {75}{2007} {053406}.
\bibitem{bichromaticonemli3} \Name{S. Flach and S. Denisov} Acta \REVIEW{ Phys. Pol.} {35}{2004} {1437}.
\bibitem{bichromatic2} \Name{M. V. Fistul, A. E. Miroshnichenko and S. Flach} \REVIEW{ Phys. Rev. B} {68} {2003} {153107}.
\bibitem{bichromatic3} \Name{Lukasz Machura, Jerzy Luczka} \REVIEW{ Phys. Rev. E} {82} {2010} {031133}.
\bibitem{bichromatic4} \Name{A. B. Kolton and F. Renzoni} \REVIEW{ Phys. Rev. A} {81} {2010} {013416}.
\bibitem{bichromatic5} \Name{M. Brown and F. Renzoni} \REVIEW{ Phys. Rev. A} {77} {2008}{033405}.
\bibitem{bichromatic6} \Name{S. Denisov, L. Morales-Molina, S. Flach, P. Hanggi} \REVIEW{ Phys. Rev. A} {75} {2007}{063424}.
\bibitem{bichromatic7} \Name{Astha Sethi, Srihari Keshavamurthy} \REVIEW{ J. Chem. Phys.} {128}  {2008}{164117}.
\bibitem{bichromaticrathced1} \Name{M. Borromeo, F. Marchesoni} \REVIEW{ Phys. Rev. E} {73} {2006} {016142}.
\bibitem{ekref} \Name{C. Cedzich, T. Rybar, A. H. Werner, A. Alberti, M. Genske, R. F. Werner}  \REVIEW{ arXiv:1302.2081} {} {}.
\bibitem{peierls1} \Name{J. Struck C. Olschlager, M. Weinberg P. Hauke J. Simonet A. Eckardt M. Lewenstein K. Sengstock and P. Windpassinger} \REVIEW{ Phys. Rev. Lett.} {108} 225304 {2012}.
\bibitem{peierls2} \Name{Krzysztof Sacha Katarzyna Targonska and Jakub Zakrzewski} \REVIEW{ Phys. Rev. A} {85}
{2012} {053613}.
\bibitem{thommen2} \Name{Quentin Thommen, Jean Claude Garreau and Veronique Zehnle} \REVIEW{ Phys. Rev. A} {84}
{2011} { 043403}.
\end{thebibliography}
\end{document}